\documentclass[aps,prb,reprint,groupedaddress,amsmath,showpacs]{revtex4-1}

 \usepackage{graphicx}
\usepackage[
    bookmarks,
    bookmarksopen=true,
    colorlinks=true,
    linkcolor=red, 
    anchorcolor=black,
    citecolor=blue, 
    filecolor=cyan, 
    menucolor=red, 
    urlcolor=blue,
    plainpages=false, 
    pdfpagelabels, 
    hypertexnames=false, 
    linktocpage 
]{hyperref}

\newcommand{\abs}[1]{\left| #1 \right|} 
 
\newcommand{\ket}[1]{\left| #1 \right>} 
\newcommand{\braket}[2]{\left< #1 \vphantom{#2} \right|
 \left. #2 \vphantom{#1} \right>} 
\let\baraccent=\= 
\renewcommand{\=}[1]{\stackrel{#1}{=}} 
\newcount\colveccount
\newcommand*\colvec[1]{
        \global\colveccount#1
        \begin{pmatrix}
        \colvecnext
}
\def\colvecnext#1{
        #1
        \global\advance\colveccount-1
        \ifnum\colveccount>0
                \\
                \expandafter\colvecnext
        \else
                \end{pmatrix}
        \fi
}

\begin{document}


\title{Emission of time-bin entangled particles into helical edge states}


\author{Patrick P. Hofer}
\email[]{patrick.hofer@unige.ch}
\author{Markus B\"uttiker}
\affiliation{D\'epartement de Physique Th\'eorique, Universit\'e de Gen\`eve, CH-1211 Gen\`eve, Switzerland}


\date{\today}

\begin{abstract}
We propose a single-particle source which emits into the helical edge states of a two-dimensional quantum spin Hall insulator. Without breaking time-reversal symmetry, this source acts like a pair of noiseless single-electron emitters which each inject separately into a chiral edge state. By locally breaking time-reversal symmetry, the source becomes a proper single-particle emitter which exhibits shot noise. Due to its intrinsic helicity, this system can be used to produce time-bin entangled pairs of electrons in a controlled manner. The noise created by the source contains information on the emitted wavepackets and is proportional to the concurrence of the emitted state.
\end{abstract}


\maketitle



\textit{Introduction}. Control over quantum-coherent electron transport on the single-particle level \cite{moskalets:book} promises benefits in a variety of research areas ranging from quantum computation \cite{nielsen:2000} to quantum metrology \cite{pekola:2008}.
In this spirit, much research was focused on electronic few-particle processes, probing the statistics of the involved charge carriers \cite{bocquillon:2013rev,mahe:2010,albert:2010,bocquillon:2012}. The required electron waveguides are usually provided by the chiral edge states which arise in the quantum Hall regime \cite{buttiker:1988}.
Devices of particular interest are synchronized emitters which permit the on-demand creation of coherent few-particle states \cite{olkhovskaya:2008,splettstoesser:2009,bocquillon:2013}.
Different means of realizing a single-electron source (SES) were investigated theoretically \cite{moskalets:2008,keeling:2006,keeling:2008}, as well as experimentally \cite{feve:2007,connolly:2013,fricke:2013,pothier:1992,dubois:2013}, notably the SES provided by a mesoscopic capacitor \cite{moskalets:2008,keeling:2008,feve:2007}.

\begin{figure}
\includegraphics[width=.85\columnwidth]{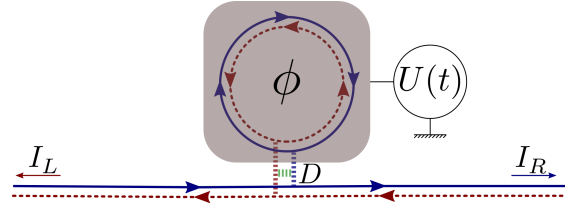}%
\caption{(color online). Helical single-electron source. A quantum dot is tunnel coupled to helical edge states through a quantum point contact of transmission $D$. A top gate (grey shading) is used to vary the potential $U(t)$ and thereby shift the discrete levels of the quantum dot. Every time an energy level is shifted above (below) the Fermi energy, an electron (hole) is emitted into the edge states. The current that leaves the source to the right (left) is labeled $I_R$ ($I_L$) and $\phi$ denotes a magnetic flux that threads the quantum dot. \label{fig:heses}}
\end{figure}

In this Rapid Communication, we propose an analogous SES in a system where the waveguides are provided by the helical edge states given in a two-dimensional quantum spin Hall insulator, a topologically nontrivial state of matter which recently received much attention \cite{kane:2005,bernevig:2006,konig:2007,du:2013}. The helicity refers to the fact that particles related by time-reversal symmetry (TRS) occupy different channels which propagate in opposite directions (cf.$\:$Fig.$\:$\ref{fig:heses}). These channels are topologically protected from backscattering as long as TRS is preserved.

An immediate consequence of substituting chiral with helical waveguides is the involvement of TRS-related partners termed Kramers pairs. This allows us to investigate entanglement which is in itself an intriguing manifestation of nonlocality in quantum mechanics, as well as a resource for quantum computation \cite{nielsen:2000}.
In this Rapid Communication, we show that the proposed SES, due to its intrinsic helicity, can be used to create time-bin entanglement \cite{brendel:1999,chirolli:2011} between two spatially separated parties \cite{wiseman:2003}. Our proposal exploits helical edge states not only for detection \cite{sato:2010,chen:2012}, but for creation of entanglement \cite{sato:2013}. The zero temperature shot noise created by the proposed source provides a measure for this entanglement.

\textit{Single-electron source}. The proposed SES, consisting of a quantum dot (QD) tunnel coupled to extended edge states by a quantum point contact (QPC), is sketched in Fig.$\:$\ref{fig:heses}.
In contrast to the extended edge states, which are described by a continuous, linear dispersion relation, the QD provides discrete edge state levels. A top gate can be used to shift these energy levels with respect to the Fermi energy, $E_F$, defined by the population of the extended edge states. Every time an energy level crosses $E_F$, an electron is emitted or absorbed by the QD, depending on whether the level moves above or below $E_F$. If TRS is preserved, the levels in the QD come in Kramers pairs which cross $E_F$ at equal times. Therefore, a particle is emitted into each channel at these times, a scenario that corresponds to two copies of the analogous chiral SES \cite{moskalets:2008, keeling:2008}.
A qualitatively different source is obtained by threading the QD with an Aharonov-Bohm (AB) flux which locally breaks TRS, lifting the Kramers degeneracy. Each level then crosses $E_F$ at a different time. Due to spin-orbit coupling, the particles have a finite probability of flipping their spin in the emission process \cite{krueckl:2011,delplace:2012}, which corresponds to a change in the propagation direction. The emitted particles are thus in a superposition of left- and rightmovers, leading to entanglement as discussed below.

For a quantitative investigation of this device, we resort to the non-interacting (Floquet) scattering matrix approach \cite{moskalets:2002}, which was shown to be a good approximation for helical edge modes in quantum spin Hall systems \cite{teo:2009}. The frozen (in time) scattering matrix of the SES can be calculated as in Ref.$\:$\onlinecite{delplace:2012} and reads (up to a global phase)
\begin{equation}
\label{eq:sgen}
 S=\begin{pmatrix}
   Y_{-} & -d^*d_{\sigma}\bar Z \\
   -dd_{\sigma}^*\bar Z & Y_+
 \end{pmatrix},
\end{equation}
where \:$Y_{\pm}=-\abs{r}+\abs{d}^2Z_\pm+\abs{d_{\sigma}}^2Z_\mp$,
\begin{equation}
\label{eq:z}
Z_{\pm}=\frac{e^{i(\varphi\pm\phi)}}{1-\abs{r}e^{i(\varphi\pm\phi)}},
\end{equation}
 and $\bar Z=Z_+-Z_-$. The amplitudes $r$, $d$, and $d_\sigma$ denote reflection and transmission at the QPC, $\sigma$ labels a spin flip (change of propagation direction), and $D=\abs{d}^2+\abs{d_{\sigma}}^2=1-\abs{r}^2$ is the total transmission probability. The phase picked up during one round trip in the QD consists of the AB phase $\phi$ and the dynamical phase $\varphi=2\pi[E-U(t)]/\Delta$, where $\Delta$ is the level spacing of the QD, $U(t)$ the time-dependent potential applied by the top gate and $E$ the energy of an incoming particle.
 
 We first discuss the TRS preserving case of $\phi=0$. In agreement with the absence of backscattering, $S$ is diagonal in this case. The diagonal elements correspond to the scattering matrix of a chiral SES and they are identical, due to the Kramers degeneracy. 
 
In order to derive analytic results, we consider the adiabatic limit. There the scattering matrix changes over timescales much larger than the dwell time $\tau=h/(\Delta D)$, the time an electron stays in the cavity. For small transmission $D$, the energy levels in the QD are well resolved and the scattering matrix only deviates significantly from unity around resonances ($\varphi=0$). We therefore expand $S$ up to linear order in $D$ and $\varphi$. Anticipating that the driving potential, $U(t)$, is well approximated as linear in $t$ around the relevant resonances, we expand $\varphi$ up to linear order in $t$. The diagonal entries of $S$ then read \cite{olkhovskaya:2008}
 \begin{equation}
 \label{eq:sc}
   S^c=\frac{t-t^E+i\Gamma}{t-t^E-i\Gamma},
 \end{equation}
where the superscript $c$ denotes the equivalence to the chiral SES.
The time of resonance at a particular energy is determined by $U(t^E)=E$ and its duration is characterized by
\begin{equation}
\label{eq:gamma}
\Gamma=\frac{D\Delta}{4\pi}\left(\frac{dU}{dt}\right)^{-1}\bigg|_{t=t^E}.
\end{equation}
Although $\Gamma$ is in principle dependent on energy, we omit an additional superscript $E$.

\textit{Current}. At zero temperature, the emitted current can be evaluated by \cite{buttiker:1994}
\begin{equation}
  \label{eq:curr}
  I_\alpha(t)=-\frac{ie}{2\pi}\left[S(E_F)\frac{\partial S^\dag(E_F)}{\partial t}\right]_{\alpha\alpha},
\end{equation} 
where $\alpha$ stands for the scattering channel. For the single-channel scattering matrix $S^c$, this leads to a Lorentzian shaped current pulse every time an energy level crosses $E_F$ \cite{olkhovskaya:2008}
\begin{equation}
  \label{eq:lorentz}
  I(t)=\frac{e}{\pi}\frac{\Gamma}{\left(t-t^{E_F}\right)^2+\Gamma^2},
\end{equation}
where $\Gamma$ is evaluated at $t^{E_F}$.
 The integral over one pulse corresponds to a single elementary charge and the sign of the current is determined by whether the level moves above or below the Fermi energy [cf.$\:$Eq.$\:$\eqref{eq:gamma}].
 Preserving TRS, the proposed helical SES therefore simultaneously emits left- and rightmoving electrons or holes in a controlled manner, acting like two copies of the chiral SES.

We next ask whether this source can emit particles that are entangled. To this extend we turn to the case of a finite AB flux that threads the QD and locally breaks TRS. This lifts the Kramers degeneracy, such that each energy level crosses $E_F$ at a different time.
For a splitting that is smaller than the level spacing, the magnetic flux needs to be smaller than one flux quantum. For a QD diameter of $1\:\mu m$ \cite{feve:2007}, this indicates a magnetic field of the order of $1\:$Gauss ($10^{-4}\:$Teslas).

Diagonalizing Eq.$\:$\eqref{eq:sgen}, the scattering eigenchannels (labelled $\pm$) are
\begin{equation}
  \label{eq:sdiag}
   S_d=V^{\dag}SV\approx\text{diag}\{S^c_-,\:S^c_+\},\hspace{.3cm}
    V=\frac{1}{\sqrt{D}}
    \begin{pmatrix}
      \abs{d} & -\frac{d_\sigma d^*}{\abs{d}}\\
      \frac{d_\sigma^*d}{\abs{d}} & \abs{d}
\end{pmatrix}.
\end{equation}
The entries are again of the chiral form and can be approximated by Eq.$\:$\eqref{eq:sc} with different emission times and widths determined by $U(t^E_\pm)=E\pm\Delta\phi/2\pi$ and Eq.$\:$\eqref{eq:gamma} evaluated at $t^{E}_\pm$. The current in the eigenchannels of the scattering matrix, $I_\pm$, is therefore given by Eq.$\:$\eqref{eq:lorentz} with the substitution $t^{E_F}\rightarrow t_\pm=t^{E_F}_\pm$ and $\Gamma\rightarrow\Gamma_\pm=\Gamma(t^{E_F}_\pm)$. The current in the left- and rightmoving channels can then easily be obtained using Eq.$\:$\eqref{eq:curr}.
Around the time a rightmover crosses the Fermi energy it reads
\begin{equation}
  \label{eq:curr2}
  \boldsymbol{I}(t)=\colvec{2}{I_R}{I_L}=\frac{e}{\pi D}\colvec{2}{\abs{d}^2}{\abs{d_\sigma}^2}\frac{\Gamma_-}{\left(t-t_-\right)^2+\Gamma_-^2}.
\end{equation}
The current emitted by a leftmover crossing the Fermi energy is obtained by substituting $d\leftrightarrow d_\sigma$ and the index $-\rightarrow+$.

\begin{figure}
\includegraphics[width=.85\columnwidth]{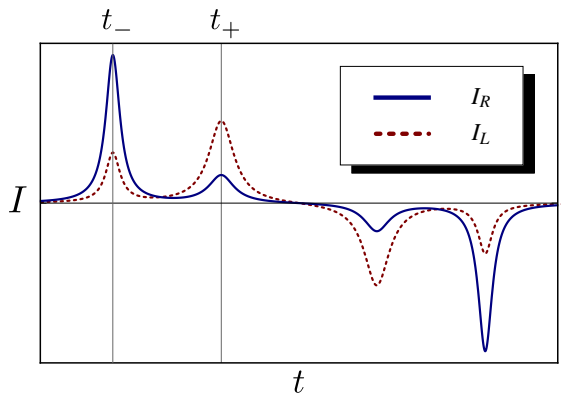}%
\caption{(color online). Current emitted by a helical single-electron source in the presence of an Aharonov-Bohm flux as a function of time (arbitrary scale). The blue (solid) line shows the excess current that flows to the right, red (dashed) denotes the leftmoving current. At each of the four times an energy level crosses the Fermi energy, an electron or hole is emitted in a superposition of rightmover and leftmover. Parameters: $U_0=-\Delta/8$, $U_1=\Delta/4$, $|d|^2=0.15$, $|d_\sigma|^2=0.05$, $\phi=\pi/6$.\label{fig:hecurr}}
\end{figure}

Applying an oscillating top gate potential which moves one (split) Kramers pair above and below $E_F$, the SES can emit single particles and holes in a periodic manner. 
The current that is created by the driving potential $U(t)=U_0+U_1\cos(\Omega t)$ is plotted in Fig.$\:$\ref{fig:hecurr}.
The peaks correspond to energy levels moving above $E_F$, emitting electrons, the dips denote hole emissions. The relative height of the peaks in the different channels is solely determined by the ratio of $\abs{d}^2$ to $\abs{d_\sigma}^2$ [cf.$\:$Eq.$\:$\eqref{eq:curr2}] and the integral over the sum of the currents corresponds to exactly one electron (hole) per peak (dip). By locally breaking TRS, the proposed source therefore becomes a proper single particle emitter. Allowing for spin flips in the emission process, the emitted particles are in a superposition of right- and leftmovers.

\textit{Noise}. We continue the characterization of the source by calculating its zero frequency shot noise at zero temperature \cite{olkhovskaya:2008}.
In the case of preserved TRS, the two emitted particles are forced into different outputs by the Pauli principle, leading to a \textit{noiseless} source. If TRS is broken, the probability of emitting two particles into the same channel and therefore also the noise become finite.

The autocorrelated zero frequency noise can be written as \cite{olkhovskaya:2008}
\begin{equation}
  \label{eq:noise}
\mathcal{P}=\frac{e^2\Omega}{2\pi}\sum_{q=-\infty}^\infty\abs{q}\left[\abs{\left\{\abs{S_{RL}}^2\right\}_q}^2+\abs{\left\{S_{RR}^*S_{RL}\right\}_q}^2\right],
\end{equation}
where $\{\cdots\}_q$ denotes the $q$th Fourier coefficient and $S_{\beta\alpha}$ is the matrix element of Eq.$\:$\ref{eq:sgen} which relates the incoming channel $\alpha$ to the outgoing channel $\beta$.
Evaluating the above expression to lowest order in $\Omega$ leads to $\mathcal{P}=\mathcal{P}^e+\mathcal{P}^h$, where
\begin{equation}
  \label{eq:helicalnoise}
  \mathcal{P}^i=\frac{e^2\Omega}{2\pi}\frac{4\abs{d}^2\abs{d_\sigma}^2}{D^2}\left\{1-\frac{4\Gamma^i_-\Gamma^i_+}{\left[\left(\Gamma^i_++\Gamma^i_-\right)^2+\bar{t}_i^2\right]}\right\}.
\end{equation}
Here $i=e/h$ labels electron or hole emission and $\bar{t}=t_+-t_-$. 
The noise vanishes as soon as there is no uncertainty in how many particles propagate to which side. This happens in the TRS preserving case ($\Gamma_-=\Gamma_+$ and $\bar{t}=0$) and if $\abs{d}$ or $\abs{d_\sigma}$ is equal to zero. In both cases one particle is emitted into each direction.

Equation \eqref{eq:helicalnoise} is equivalent to the noise generated by two chiral sources, which have their outgoing channels coupled by a QPC with transmission and reflection probabilities $\abs{d}^2/D$, $\abs{d_\sigma}^2/D$ \cite{olkhovskaya:2008,bocquillon:2013}.
We therefore interpret the noise as an interplay of exchange and partition noise \cite{blanter:2000}, equivalent to a Hong-Ou-Mandel (HOM) setup \cite{hong:1987} where two Fermions are sent to a beamsplitter with a time difference \cite{olkhovskaya:2008,bocquillon:2013}. In our case the beamsplitter is given by the QPC, through which the source emits two particles with time difference $\bar{t}$.

\begin{figure}
\includegraphics[width=.95\columnwidth]{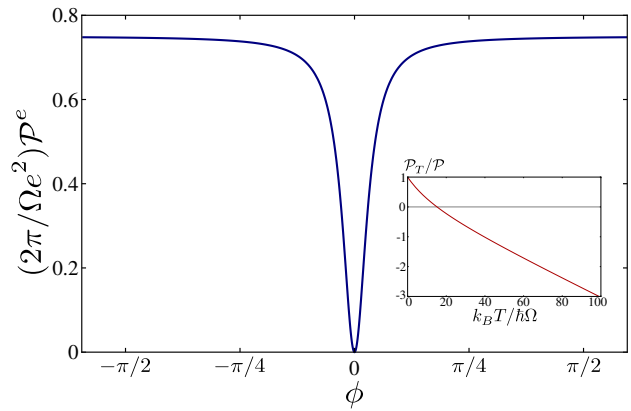}%
\caption{(color online). Zero temperature shot noise created by the source as a function of the magnetic flux. At zero flux, equality of the emission times forces the particles into different outputs. For finite flux, the emission times differ and the noise increases until it saturates when the emitted particles have no overlap anymore. The noise is proportional to the concurrence created per cycle. The inset shows the excess noise divided by the zero temperature noise as a function of temperature. Here: $|d|^2=0.15$, $|d_\sigma|^2=0.05$.\label{fig:noise}}
\end{figure}

If the top gate potential is linear throughout the emission of Kramers pairs, i.e. $\Gamma_-=\Gamma_+$ for electrons and holes respectively, the noise generated by the electrons is equal to the noise generated by the holes and can be expressed solely as a function of the AB phase and the transmission probabilities
\begin{equation}
  \label{eq:helicalnoiselin}
  \mathcal{P}^e=\mathcal{P}^h=\frac{e^2\Omega}{2\pi}\frac{4\abs{d}^2\abs{d_\sigma}^2}{D^2}\frac{\phi^2}{(D/2)^2+\phi^2}.
\end{equation} 
Note that electron and hole emissions can happen with different velocities.
The noise created by electron emission is plotted in Fig.$\:$\ref{fig:noise}.

\textit{Entanglement}. To investigate entanglement, we consider the case where one Kramers pair is linearly moved above $E_F$, $U=\Delta\Omega t$.
 Using an undisturbed Fermi sea, denoted $\ket{0}$, as the incoming state leads to the outgoing state \cite{keeling:2006,keeling:2008}
\begin{equation}
  \label{eq:out2}
  \ket{\Psi_\text{out}}=B^\dag_+B^\dag_-\ket{0},\hspace{.2cm}B_\pm=\sqrt{\frac{2{\Gamma}}{\hbar}}\int_0^\infty dEe^{\frac{E}{\hbar}(it_\pm-{\Gamma})}b_\pm(E).
\end{equation}
This state consists of a single electron in each scattering eigenchannel on top of an undisturbed Fermi sea.

We now consider the entanglement between two spatially separated parties, who respectively receive the left- or rightmoving particles \cite{wiseman:2003}.
The last state consists of terms where both electrons propagate in the same direction in addition to terms where one left- and one rightmover is emitted. Since the particle number is separately conserved for each channel, the subspaces of different particle numbers in a given channel have to be treated separately and their respective entanglement summed up \cite{wiseman:2003,beenakker:2006}. Obviously the states where two particles propagate in the same direction are not entangled. The remaining part of the state reads
\begin{equation}
\label{eq:time}
  \ket{\Psi_{LR}}=\frac{1}{D}\left(\abs{d}^2\ket{+_L,-_R}+\abs{d_\sigma}^2\ket{-_L,+_R}\right),\hspace{.2cm}
\end{equation}
where $\ket{\pm}_{L/R}=B^\dag_{L/R\pm}\ket{0}$ describes an electron emitted at time $t_\pm$ into channel $L/R$ and the operator $B_{L/R\pm}$ is obtained by substituting $b_\pm\rightarrow b_{L/R}$ in $B_\pm$, cf.$\:$Eq.$\:$\eqref{eq:out2}. The normalization is chosen such that the absolute square gives the probability of finding an electron in each channel. The entanglement of the last state can be quantified by the concurrence \cite{wootters:1998} emitted per cycle \cite{beenakker:2005}, which is the total concurrence times the probability of finding this state
\begin{equation}
  \label{eq:concurrence}
  \mathcal{C}=\frac{2\abs{d}^2\abs{d_\sigma}^2}{D^2}\left(1-\abs{\braket{-}{+}}^2\right).
\end{equation}
The overlap integral $\abs{\braket{-}{+}}^2=4\Gamma_-\Gamma_+/[(\Gamma_-+\Gamma_+)^2+\bar{t}^2]$ is equal for leftmovers and rightmovers.
The last equation reaches the theoretical maximum of $\mathcal{C}=1/2$ \cite{samuelsson:2005,beenakker:2005} for $\abs{d}=\abs{d_\sigma}$ and well separated emission times, such that $\braket{-}{+}=0$. In that case, Eq.$\:$\eqref{eq:time} describes a Bell state which is found with probability one half.

Note that an analogous discussion holds for holes when moving a Kramers level pair below $E_F$.
Under optimal conditions, the proposed source can thus emit pairs of time-bin entangled particles on demand with the maximal production rate of half a Bell pair per cycle. 

The concurrence per cycle is equal to the created noise up to a constant factor, $(\pi/2\Omega e^2)\mathcal{P}=\mathcal{C}$, providing a simple connection between a measurable quantity and the created entanglement. 
Since the two quantities have their origin in different properties of the emitted state, this relation is not due to a fundamental connection of entanglement and noise, but simply a result of current conservation.
The concurrence results from the subspace where one particle propagates in each direction, whereas the noise is due to the particle number uncertainty in each channel.
However, being able to write the noise in terms of the overlap of the emitted particles, i.e. in the form of Eq.$\:$\ref{eq:concurrence}, supports the analogy to a HOM experiment \cite{blanter:2000} and directly connects the noise to the emitted wavepackets.

\textit{Finite temperature}. All the considerations in this section hold as long as the temperature is smaller than the energy scale over which the scattering matrix varies. In our case this is determined by the level width. For a level spacing $\Delta\approx2.5\:K$ (as in the experiment of Ref.\:\onlinecite{feve:2007}), this implies $k_BT\ll500\:mK$.
In this regime, the current does not change but the noise is modified and acquires a $\phi$-independent background.
The excess noise $\mathcal{P}_T$ is defined as the total noise minus the noise measured when the source is off and the magnetic flux is zero. This quantity is proportional to the zero temperature shot noise at all temperatures (see inset of Fig.$\:$\ref{fig:noise}). The shape of the excess noise is therefore temperature independent.

Previous work \cite{samuelsson:2009} indicates that the entanglement only survives up to temperatures of the order of the driving frequency. Experimentally, the adiabatic regime is therefore more challenging than the non-adiabatic regime, where the emitted wavepackets have a different form \cite{keeling:2008} and our results are only qualitatively valid. However, since the shape of the excess noise as well as the concurrence depend only on the overlap of the emitted particles, they are proportional to each other also in this regime.

\textit{Conclusions}. We propose a SES which emits particles into helical edge states. If TRS is preserved, this source equals two copies of the known, noiseless chiral SES \cite{moskalets:2008, feve:2007}. Locally breaking TRS by applying an AB flux leads to proper single-particle emission. Due to a finite spin flip probability in the emission process, the emitted particles are in a superposition state of left- and rightmovers and the source becomes noisy. The resulting current correlation can be accounted for by an interplay of exchange and partition noise generated at the QPC through which the source emits, analogous to a HOM experiment. It contains information on the emitted wavepackets and only depends on the applied AB flux and the transmission probabilities.

Due to the intrinsic helicity, the particles are emitted in a superposition of states which are separated in real space. This leads to time-bin entanglement when emitting single Kramers pairs. The concurrence created per cycle is proportional to the noise, relating it to an experimentally available quantity. Under optimal conditions, the concurrence reaches the theoretical maximum which corresponds to the emission of half a Bell pair per cycle.
Using the setup of Ref.$\:$\onlinecite{splettstoesser:2009}, this entanglement could in principle be used to maximally violate a Bell inequality.

Taking into account Coulomb interactions could lead to interesting corrections to the results and possibly also to additional applications for a mesoscopic capacitor in the TI regime.
Although the calculations are done in the adiabatic regime, the results should qualitatively be the same in the experimentally relevant non-adiabatic regime, where finite temperatures have little effect.

\textit{Acknowledgements}. During the preparation of this manuscript we learned about related work \cite{inhofer:2013,ferraro:2013}.

We acknowledge guidance through the review process by Christian Flindt and useful discussions with Michael Moskalets, Jian Li, David Dasenbrook, and Christoph Schenke.
This work was funded by the SNF and the NCCR QSIT.

\end{document}